\documentclass[prl,reprint,showpacs,twocolumn,superscriptaddress,floatfix]{revtex4-1}
\usepackage{epsfig}
\usepackage{amsmath}
\usepackage{amssymb}
\usepackage{amsfonts}
\usepackage{mathptmx}
\usepackage{dcolumn}
\usepackage{eucal}
\usepackage{bm}
\usepackage{color}
\usepackage[colorlinks,linkcolor=blue,citecolor=blue]{hyperref}
\usepackage{epstopdf}
\usepackage{bbold}
\usepackage{url}

\usepackage{dsfont}

\newcommand{\Time}{\mathcal{T}}

\newcommand{\average}[1]{\langle #1 \rangle}
\newcommand{\Trace}{{\rm Tr}}

\newcommand{\hata}{\hat{a}}

\def \ket#1{\mathinner{|{#1}\rangle}}
\def \bra#1{\mathinner{\langle{#1}|}}

\def\braket#1{\mathinner{\langle{#1}\rangle}}

\newcommand\mean[1]{\mathinner{\langle{#1}\rangle}}

\newcommand{\ketbra}[2]{{\mathinner{| {#1} \rangle \langle {#2} |}} }

\newcommand{\matrixel}[3]{{\mathinner{\langle{#1}| {#2} | {#3}\rangle}} }

\newcommand{\Forder}{\overrightarrow{T}}

\newcommand{\GFunc}{\mathcal{G}}
\newcommand{\Prob}{\mathcal{P}}

\newcommand{\Real}{\Re{\rm e}}

\newcommand{\rhocoh}{\rho^0_{\rm coh}}
\newcommand{\rhoin}{\rho^0_{\rm in}}
 
\begin{document}

\title{Probing Quantum Interference Effects in the Work Distribution}

\author{P. Solinas}
\affiliation{SPIN-CNR, Via Dodecaneso 33, 16146 Genova, Italy}
\author{S. Gasparinetti}
\affiliation{Department of Physics, ETH Z\"urich, CH-8093 Z\"urich, Switzerland}

\date{\today}

\begin{abstract}
What is the role of coherence in determining the distribution of work done on a quantum system? We approach this question from an operational perspective and consider a setup in which the internal energy of a closed system is recorded by a quantum detector before and after the system is acted upon by an external drive. 
We find that the resulting work distribution depends on the initial state of the detector as well as on the choice of the final measurement. We consider two complementary measurement schemes, both of which show clear signatures of quantum interference.
We specifically discuss how to implement these schemes in the circuit QED architecture, using an artificial atom as the system and a quantized mode of the electromagnetic field as the detector. Different measurement schemes can be realized by preparing the field either in a superposition of Fock states or in a coherent state and exploiting state-of-the art techniques for the characterization of microwave radiation at the quantum level. More generally, the single bosonic mode we utilize is arguably the minimal quantum detector capable of capturing the complementary aspects of the work distribution discussed here.
\end{abstract}

\maketitle

\section{Introduction}

Thermodynamics, a cornerstone of modern physics, only applies to macroscopic systems. As the dimensions of the system under consideration shrink down to the atomic scale, one has to account for the increasing role played by fluctuations \cite{Seifert2012}. Furthermore, as microscopic systems obey the laws of quantum mechanics, quantum effects are expected to play an increasingly important role. Whereas it has long being theoretically established that quantum computation and cryptography can outperform their classical counterparts \cite{nielsen-chuang_book}, the question whether quantum mechanics is a true game changer in the thermodynamics arena stands as an open one in the field. 
Answering this question in the affirmative may have a tremendous impact in the way we efficiently generate, store, and distribute energy. There are strong indications that fundamental biological processes such as photosynthesis can be explained only by quantum mechanics. In particular, high efficient energy transport seems to need quantum coherent and non-Markovian dynamics \cite{Engel2007, Collini2010, Panitchayangkoon2010, mohseni2008environment, Lloyd2011}. Another significant example comes from the study of quantum heat engines \cite{Kosloff-Levy2014}: quantum coherence cannot be exploited to beat the Carnot efficiency in extracting energy from thermal baths, yet it does make it possible to boost the power output \cite{Alicki1979,Uzdin2015}.

In order to study the thermodynamics and statistical mechanics of quantum systems, one is first required to find suitable extensions of the concept of mechanical work. The latter is customarily defined as the integral of the force applied by an external agent times the displacement of the system.
Applying this definition to the quantum domain presents some challenges, even when restricting one's attention to closed systems.
Since work cannot be represented by a Hermitian operator \cite{talkner2007fluctuation}, it is not an observable in the standard quantum mechanical sense. Therefore, in order to define work one has to also specify a suitable scheme to measure it \cite{solinas2015fulldistribution}. 
A favored choice in the literature is based on a double projective measurement of the Hamiltonian \cite{kurchan2000quantum, tasaki2000jarzynski, engel2007jarzynski, campisi2011colloquium,campisi2011erratum, esposito2009Erratum,esposito2009nonequilibrium, roncaglia2014work, gasparinetti2014heat, Jarzynski2015, Carrega2016Energy}. Based on this definition, one recovers a quantum-mechanical version of the fluctuation relations originally derived in the framework of classical stochastic thermodynamics \cite{campisi2011colloquium}. These relations, which describe quantum evolution starting from a thermal state, were recently tested in experiments \cite{Batalho2014,an2014experimental}.
Still, one may speculate whether initial coherences -- and quantum interference in general -- play any role in determining the statistics of work. This question clearly cannot be addressed by a double projective measurement, as the first measurement would erase any initial coherence in the energy eigenbasis and thereby rule out the possibility that ``interfering alternatives'' \cite{feynman1965quantum, Cotler2016} leave a trace in the work distribution.

Here we consider a measurement setup for the work distribution, based on dispersive coupling to a quantum detector at the beginning and at the end of the evolution. We discuss two measurement schemes -- denoted ``A'' and ``B'' in the following -- that differ for the initial state of the detector as well as for the choice of the final measurement. Scheme A naturally leads to the definition of a quasiprobability distribution for the work, whose negativity is due to quantum coherence of the initial state and quantum interference effects \cite{Leggett1985, clerk2011full, bednorz2012nonclassical, hofer2016, Perarnau-Llobet2016}. In scheme B, quantum interference effects still appear, but only as the result of an incomplete measurement. Finally, we put forward a realization of our setup in a circuit quantum electrodynamics (circuit-QED) architecture \cite{blais2004cavity, Wallraff2004, koch2007charge}, using a transmon qubit as the system and a quantized cavity mode as the detector. Our proposed realization, which we claim to be within reach of state-of-the-art superconducting circuit technology, also illustrates the use of a single bosonic mode to probe both classical and nonclassical features of the work distribution by by preparing the mode either in a classical or in a nonclassical state.

While unifying and extending previous work, our approach highlights two complementary aspects of the work distribution that can be unveiled one at a time, but not simultaneously, in according to general principles of quantum mechanics. On one hand, our scheme A is similar in spirit to the interferometric schemes described in \cite{mazzola2013measuring,dorner2013extracting, Batalho2014}. On the other hand, the idea of directly extracting the work distribution out of a single measurement (as done here in scheme B) can already be found in \cite{roncaglia2014work,DeChiara2015,an2014experimental}. Notice, however, that the effect of initial coherences was not considered in these works. Finally, in \cite{talkner2016} the authors consider direct measurements of variable strength on generic initial states and reach similar results as we do for scheme B. However, they do not consider the possibility of directly extracting the moments of the work distribution using scheme A. This leads to the impossibility to measure higher-order moments (starting from the second one) in extremely weak coupling conditions.

\section{Different measurements lead to different statistics}
We consider a closed quantum system driven by an external, classical field, so that the system dynamics is described by a time-dependent Hamiltonian $H_S(t)$. We are interested in the distribution of work $W$ done by the external field between times $t=0$ and $t=\mathcal T$, or, equivalently, in the corresponding variation $\Delta U$ in the internal energy of the system.
To determine this variation, we couple the system to a quantum detector, namely, a free particle characterized by the conjugate position $x$ and momentum $p$ operators.
We assume that the particle has a large enough mass that its kinetic energy can be neglected \cite{nazarov2003full}. 
The interaction between the system and the detector is described by Hamiltonian $H_{SD}(t) =  -p ~ \beta(t) H_S(t)$.
As $H_{SD}$ commutes with $H_S$ at all times, the coupling to the detector does not induce transitions between the system eigenstates.
The time-dependent coupling strength $\beta(t)$ is chosen so that the detector ``records'' the energy of the system at the beginning and at the end of the evolution: $\beta(t)= \lambda/p_0 [ \delta(t-\Time) - \delta(t)]$, where the $\lambda$ and $p_0$ are constant parameters with the dimension of time and momentum, respectively.
The meaning of the delta functions in the expression for $\beta(t)$ is that the coupling takes place over a time scale that is fast compared to the variation of $H_S(t)$, so that the dynamics of the system is effectively ``frozen'' while the detector is coupled. Neither the system nor the detector are projectively measured between times $0$ and $\mathcal T$.

The full evolution is described by the operator
$U(\Time)= e^{i    \frac{p \lambda }{p_0}  H_S(\Time) }  U_S(\Time)   e^{-i    \frac{p \lambda }{p_0}  H_S(0)}$ \cite{solinas2015fulldistribution}, where 
$U_S(\Time) = \Forder e^{-i \int_0^\Time dt H_S(t)}$ is the evolution operator of the system alone and we have set $\hbar=1$.
We reasonably assume that the system and the detector are initially in a separable state and that the detector is prepared in a pure state. Then the initial state is described by the density matrix
$\rho(0) = \rho_S^0 \otimes \int dp dp' G^*(p') G(p) \ket{p}\bra{p'}$
, where $\rho_S^0$ is the initial density matrix of the system and $G(p)$ is the detector wavefunction in momentum space.
We denote with $\ket{\epsilon_i^t}$ the $i$-th eigenstate of $H_S(t)$ and with $\epsilon_i^t$ its corresponding eigenvalue.
The final state of the detector is described by the following density matrix, written either in the momentum basis, $\{\ket{p}\}$, or in the position basis, $\{\ket{x}\}$:
\begin{widetext}
\begin{subequations}
\begin{align}
\rho_D(\Time) &= \sum_{ikj} \int  dp dp' \rho_{ik}^0 G(p) G^*(p') U_{S,ji} U_{S,kj}^\dagger e^{i \frac{p \lambda }{p_0} \epsilon_{ji}} e^{-i \frac{p' \lambda}{p_0} \Delta \epsilon_{jk}} \ketbra{ p}{p'}
\label{eq:rhoD_momentum}\\
 &= \sum_{ikj} \int  dx dx' \rho_{ik}^0 g \left(x- \frac{\lambda}{p_0}  \epsilon_{ji}\right) g^*\left(x'-\frac{\lambda}{p_0}  \epsilon_{jk}\right) U_{S,ji} U_{S,kj}^\dagger  \ketbra{x}{x'}
\label{eq:rhoD_position}
\end{align}
  \label{eq:rho_D_p}
\end{subequations}
\end{widetext}
where $\rho_{ik}^0 = \matrixel{\epsilon_i^0}{\rho_{S}^0}{\epsilon_k^0}$, $\epsilon_{ik} = \epsilon^\Time_{i} - \epsilon^0_{k}$, 
 $g(x) = \int d p e^{- i p x} G(p)$ is the Fourier transform of $G(p)$, $U_{S,ji}=\matrixel{\epsilon_j^\Time}{U_S(\Time)}{\epsilon_i^0}$, and
$U_{S,ji}^\dagger=\matrixel{\epsilon_j^0}{U_S^\dagger(\Time)}{\epsilon_i^\Time}$.

Equations \eqref{eq:rho_D_p} are a good starting point to discuss different measurement schemes, as we have not yet specified the initial state of the detector, nor the measurement to be performed at the end.
First of all, we notice that the detector momentum $p$ is a conserved quantity of the evolution. As a result, the coupling between the system and the detector cannot induce transitions between different eigenstates of the momentum; instead, it changes their relative phase.
This picture is obviously reversed when considering eigenstates of the position, as the system-detector coupling $\beta(t)~p~H_S$ induces transition between eigenstates of the position operator $x$.
Based on these considerations, one can envisage the following two schemes to measure the work distribution:
\begin{itemize}
\item[A)] Prepare the detector in a superposition of momentum eigenstates and measure their relative phase at the end of the evolution.
	\item[B)] Prepare the detector in a position eigenstate 
	and make a position measurement.
\end{itemize}
Before discussing each scheme in more detail, we observe that scheme B is a variation of the standard von Neumann measurement scheme \cite{VonNeumann1955}, while scheme A is clearly reminiscent of the full-counting statistics formalism \cite{nazarov2003full, clerk2011full, hofer2016, Perarnau-Llobet2016}.

We first consider scheme A and focus on the phase accumulated between the momentum eigenstates $\ket{p_0/2}$ and $\ket{-p_0/2}$. This phase can be measured by standard tomography and is related to the characteristic function
$\GFunc_\lambda=  \matrixel{p_0/2 }{\rho_D(t) }{-p_0/2 }/\matrixel{p_0/2 }{\rho^0_D }{-p_0/2 }$ \cite{solinas2015fulldistribution, SM}.
The characteristic function is therefore the primary measurement outcome in scheme A. From an operational point of view, we notice that the dependence of $\GFunc_\lambda$ on $\lambda$ can be retrieved by varying either the time or the strength of the system-detector coupling.
Once $\GFunc_\lambda$ is known, the moments of the work distribution can be determined in the standard way as
\begin{equation}
\average{W^n} = (-i)^n d^n \GFunc/d \lambda^n |_{\lambda=0} \ . \label{eq:moments}
\end{equation}
For a closed system (such as the one we are considering), the moments take the intuitive form $\average{W^n} = \Trace [(U_S H_S(\Time)U_S^\dagger -  H_S(\Time))^n \rho_S(0)]$.

Can the characteristic function be linked to a probability distribution for the work? The answer is not obvious and directly points to the quantum nature of the work done on a quantum system. A distribution $\Prob(W)$ can be formally defined as the Fourier transform of the characteristic function \cite{talkner2007fluctuation}: $\Prob(W)= \int d \lambda~\exp{\{ - i  \lambda W \}} \GFunc_\lambda $. Starting from \eqref{eq:rho_D_p}, a direct calculation gives \cite{SM}
\begin{equation}
 \Prob(W) =  \sum_{ikj} \rho_{ik}^0  U_{S,ji} U_{S,kj}^\dagger \delta \Big[W - \left(\epsilon_j^\Time- \frac{\epsilon_i^0+\epsilon_k^0}{2}\right) \Big].
 \label{eq:P(W)}
\end{equation}
The diagonal contributions in \eqref{eq:P(W)}, i.e., those with $i=k$, are always positive and can be immediately interpreted in terms of classical transitions between different eigenstates (see the following and the Supplementary Material (SM) \cite{SM}).  For this reason, we refer to them as ``classical contributions''. On the contrary, the off-diagonal contributions, i.e., those with $i\neq k$, correspond to classically forbidden exchanges of energy $(\epsilon_i+\epsilon_k)/2$. Furthermore, their sign can be negative. Due to the presence of these terms, $\Prob(W)$ is not definite positive and can only be referred to as a quasi-probability distribution \cite{solinas2015fulldistribution, allahverdyan2014, hofer2016, Perarnau-Llobet2015}. The negativity of $\Prob(W)$ is a signature of the ``quantumness'' of the work distribution and can be associated to the violation of a Leggett-Garg inequality \cite{bednorz2010quasiprobabilistic,bednorz2012nonclassical, clerk2011full, solinas2015fulldistribution, Perarnau-Llobet2016}.
From the structure of Eq.~\eqref{eq:P(W)} it is clear that the presence of initial coherence between different energy eigenstates is a necessary condition for the negativity of $\Prob(W)$. This is in agreement with a recent interpretation of negative quasiprobabilities in the full counting statistics as arising from quantum interference effects \cite{hofer2016}. Notice also the analogy with the full counting statistics of charge transport in superconducting circuits \cite{Belzig2003}, with the phase coherence of the superconducting order parameter playing the role that initial-state coherences play here.

The negative quasiprobability distribution we predicted for scheme A is by no means unphysical, as it does not describe the probability of a certain measurement outcome. A more direct access to the work distribution is provided by scheme B, in which we initialize the detector in a position eigenstate and measure the distribution of its final positions. As the shift in position is proportional to the energy injected in the system, its measurement completely determines the work distribution. From Eq. (\ref{eq:rho_D_p}), the probability to find the detector in position $x$ is
\begin{equation}
  \Prob(x) = \sum_{ikj} \rho_{ik}^0  U_{S,ji} U_{S,kj}^\dagger g \left( x- \frac{\lambda \epsilon_{ji} }{p_0}  \right) g \left( x-\frac{\lambda  \epsilon_{jk}}{p_0} \right). 
  \label{eq:rho_T_gx}
\end{equation}
If the detector is initially in a definite position state, i.e., $g(x) = \delta (x-x_0)$ \cite{sokolovski2015meaning},  then only the terms with $ \epsilon_{ji} =  \epsilon_{jk}$ survive in \eqref{eq:rho_T_gx}.
Then we can write $\Prob(x-x_0) = \sum_{ij} P_{i}^0 P_{i\rightarrow j}$ where $P_{i}^0$ is the probability to find the system initially in the state $\ket{\epsilon_i^0}$ and $P_{i\rightarrow j}$ is the transition probability from state $\ket{\epsilon_i^0}$ to state $\ket{\epsilon_j^\Time}$, as prescribed by the unitary evolution $U_S$. This shows that the measurement outcome of scheme B can be fully interpreted in terms of classical conditional probabilities \cite{engel2007jarzynski}. Furthermore, $\Prob(x-x_0)$ coincides with \eqref{eq:P(W)} provided only the classical contributions are retained.

It would so far seem that no quantum interference can be observed using scheme B.
Still, interference effects are to some degree recovered if the detector cannot fully discriminate between quantum trajectories leading to different variations of the internal energy. In practice, this happens when the functions $g(x+\Delta \epsilon_{ji})$ and $g^*(x'+\Delta \epsilon_{jk})$ in Eq. \eqref{eq:rho_D_p} have a substantial overlap. Then the contributions with $\rho_{ik}^0$, $i\neq k$ do not vanish and modify the resulting distribution. This fact is fully consistent with the well-known notion of incomplete (or ``partial'') measurement in quantum mechanics \cite{feynman1965quantum}.
As we increase the uncertainty in the initial state, we continuously transition from a classical work distribution with well-resolved peaks to a distribution where individual peaks cannot be resolved. This broadened distribution is sensitive to initial coherences, thus revealing the underlying quantum dynamics and highlighting the differences between classical and quantum uncertainty.

We conclude this section by stressing one key difference between the two proposed schemes.
In scheme A, the system-detector coupling does not induce any transition between eigenstates of the detector momentum, whose relative phase is eventually measured. As a result, one gains access to all the moments of the work distribution without perturbing the dynamics, regardless of the strength of the coupling. Borrowing the terminology from circuit QED (see also experimental implementation below), we can refer to scheme A as a {\it dispersive} readout, in which the information is encoded in the phase of the detector. On the contrary, in scheme B the detector position, which is eventually measured, changes after the coupling to the system. In this case, by increasing the strength of the coupling (or, equivalently, by decreasing the uncertainty in the detector position), we partially destroy the effect of initial coherences \cite{roncaglia2014work, DeChiara2015, talkner2016}. In scheme B the perturbation introduced by the detector is unavoidable and leads to a trade-off between measurement precision and the witnessing of quantum coherence. For example, in Ref.~\cite{talkner2016} it was shown that, in the limit of very weak coupling, the moments of the work (except for the first one) are undetermined because of the uncertainty in the measurement.

\section{Implementation in circuit QED}
The concepts outlined above can be readily tested in a circuit-QED architecture \cite{Wallraff2004,blais2004cavity} using a transmon qubit \cite{koch2007charge} as the system and a quantized mode of a microwave cavity as the detector [Fig.~1(a)]. This detection scheme is not fully equivalent to that based on a free particle, introduced in the previous section, yet it preserves all its essential features provided one identifies momentum eigenstates with Fock states and position eigenstates with coherent states.
As we will see, due to the reduced dimensionality of the Hilbert space, the basis that describes the (continuous) position of our detector is overcomplete, and therefore possesses an intrinsic uncertainty. Furthermore, the kinetic energy of the detector cannot be neglected and results in an additional phase acquired by each of the Fock states \cite{SM}. We will not explicitly consider this phase as it is automatically accounted for in the standard circuit-QED measurement scheme.

In the dispersive limit, the total Hamiltonian reads
\begin{equation}
 H = \Big( \omega_a + \frac{g^2}{\Delta}\Big) \sigma_z + \omega_r \hata^\dagger \hata + \frac{g^2}{\Delta} \sigma_z \hata^\dagger \hata
 \label{eq:H_dispersive}
\end{equation}
where $\omega_a$ and $\omega_r$ are the resonant frequencies of the qubit and the cavity, respectively, $\Delta = \omega_a - \omega_r$ is the qubit-cavity detuning, $g$ is the qubit-cavity coupling,
$\hata^\dagger$ and $\hata$ are the creation and annihilation operators for the cavity mode, and $\sigma_z$ is the third Pauli operator acting on the qubit. From Equation \eqref{eq:H_dispersive} it is easy to see that the interaction Hamiltonian commutes with both the system and the detector Hamiltonians, as in the original coupling scheme.
The strength of the interaction is controlled by the detuning $\Delta$, enabling fast on/off switching. In alternative, a tunable interaction between the qubit and the cavity can be induced by an external field via higher-order transitions \cite{Gasparinetti2016}.

\begin{figure}
    \begin{center}
    \includegraphics{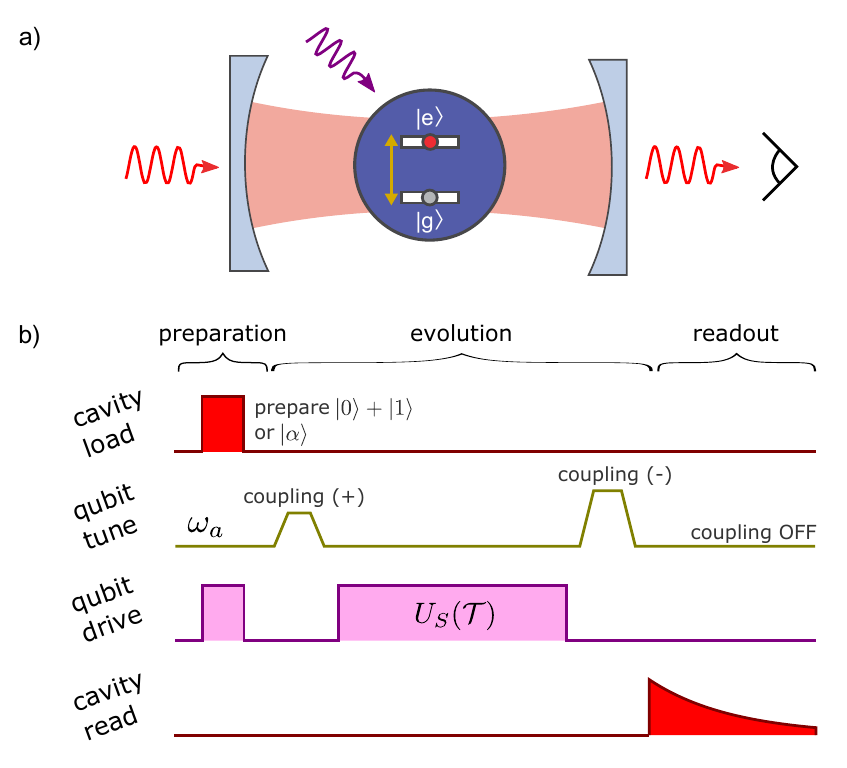}
       \end{center}
    \caption{\textbf{Measuring the work distribution in a circuit-QED architecture.} (a) Illustration of the setup. An artificial atom (the system) is dispersively coupled to a quantized field (the detector). The atom and the field can be independently addressed by control fields. The system is measured by characterizing the field leaking out of the cavity. (b) Proposed measurement scheme, applying to both scheme A and scheme B.}  
    \label{fig:cQED}
\end{figure} 

Both scheme A and scheme B can be realized by the sequence outlined in Fig.~\ref{fig:cQED} (b). It consists of a preparation (i), an evolution (ii), and a readout phase (iii). We assume that both the qubit and the cavity are initially found in the ground state.

(i) In the preparation phase, the cavity is prepared either in a superposition of two Fock states, say, $\frac{1}{\sqrt{2}}\left(\ket{0}+\ket{1}\right)$ (scheme A), or in a coherent state $\ket{\alpha}$ (scheme B). To prepare a coherent state, it is sufficient to irradiate the cavity with a coherent tone. To prepare a superposition of Fock states, we can first prepare the qubit in the desired state superposition and then transfer it to the cavity.
An additional qubit may optionally be used for this purpose.
As for the qubit, it is prepared in a pure state by applying a unitary operation.

\begin{figure}
    \begin{center}
    \includegraphics[width=\linewidth]{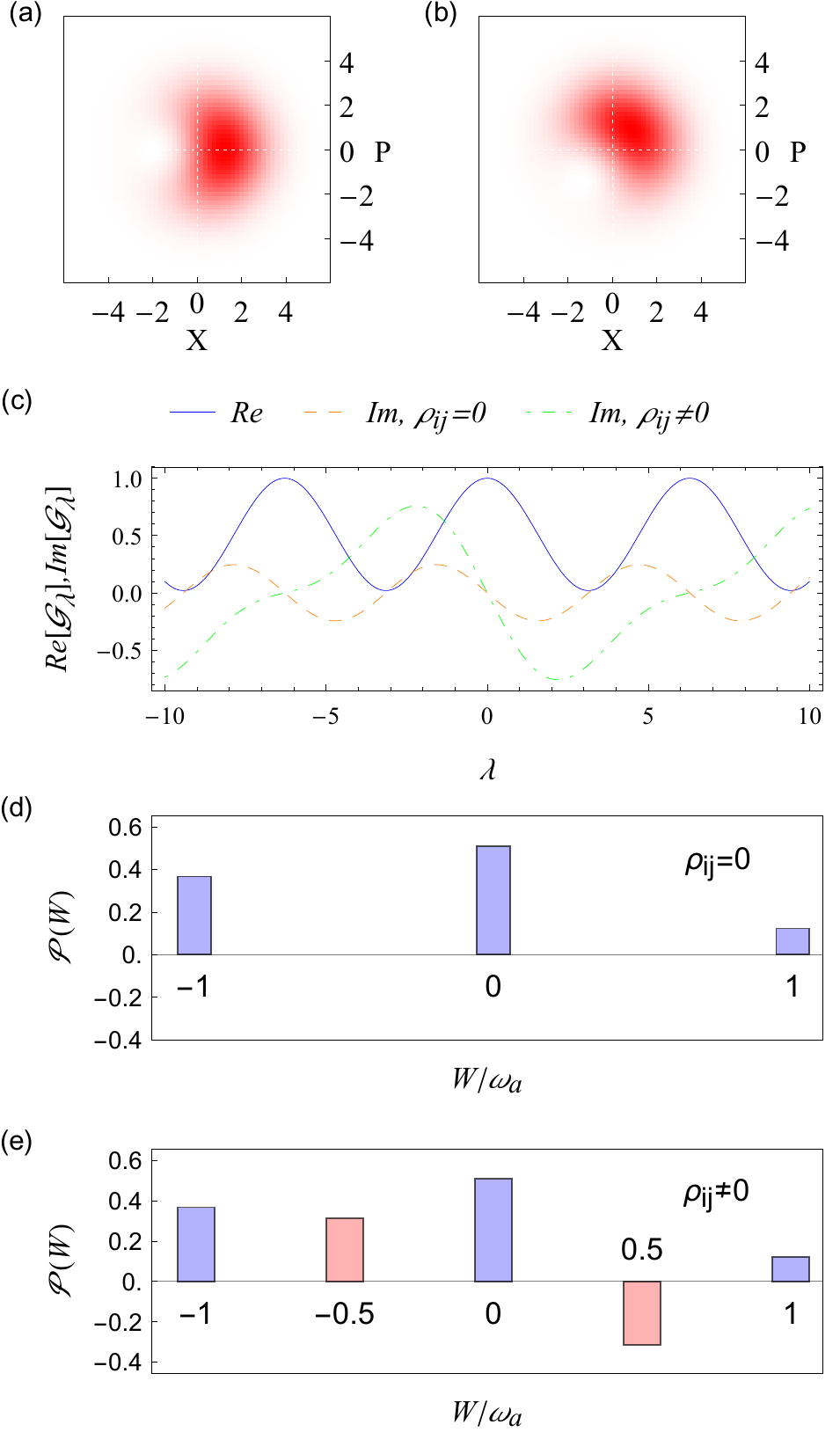}
       \end{center}
    \caption{\textbf{Results for scheme A.} (a,b) Husimi-Q function $Q_{\rho_a}(\alpha)=\langle \alpha | \rho_a | \alpha \rangle$ for the detector in the state superposition $(\ket{0}+e^{i\varphi} \ket{1})/\sqrt{2}$, with $\varphi=0$ (a) and $\varphi=0.76$ (b). (c) Real (solid) and imaginary part (dashed, dot-dashed) of the characteristic function $\GFunc_\lambda$ for an initial state with and without coherence. (d,e) Corresponding quasiprobability distributions $\Prob(W)$, as obtained from the Fourier transform of $\GFunc_\lambda$. 
    }  
    \label{fig:schemeA}
\end{figure} 

(ii) In the evolution phase, an interaction between the qubit and the cavity is induced by bringing the qubit frequency close to that of the cavity [while remaining in the dispersive regime described by \eqref{eq:H_dispersive}]. 
For $\Delta <0 $, we red-detune the system and the cavity.
This way a phase is imparted to the cavity states conditioned on the state of the qubit. Then the interaction is switched off and the desired unitary operation $U_S(\mathcal T)$ is applied to the qubit. Finally the interaction is switched on again so that the cavity can ``record'' the information on the new state of the qubit. 
To make sure the second interaction has an opposite sign to the first, we detune the the qubit and the cavity in opposite direction, i.e., use a blue-detuning.
Alternatively, one can use the same coupling scheme twice and apply a $\pi$ pulse to the qubit before and after the second coupling.

(iii) In the readout phase, the radiation leaking out of the cavity is characterized using linear amplifiers and field quadrature detection methods. The result of this characterization is a 2D histogram giving the probability that the in-phase and quadrature component of the amplified field take certain values. For a quantum-limited, phase-insensitive amplification chain, the measured histogram directly gives the Husimi-Q distribution of the field $Q_{a}(\alpha)=\frac{1}{\pi}\matrixel{\alpha}{\rho_a}{\alpha}$, where $\rho_a$ is the density matrix describing the field mode $a$ \cite{Eichler2012}. We note in passing that even in the presence of significant (thermal, uncorrelated) added noise, the measured histogram corresponds to a generalized phase space distribution, allowing for the full reconstruction (tomography) of the density matrix of the field and for the investigation of its nonclassical properties \cite{Eichler2012}.
In order to minimize the cavity leakage during the evolution phase while obtaining a good signal-to-noise ratio in the readout phase, one could make use of a tunable-coupling cavity, as demonstrated, e.g., in Ref. \cite{Yin2013}.

We will now present illustrative results for both detection schemes. We will consider two initial states: a generic superposition $\rhocoh$ of the ground and excited state, and its incoherent counterpart $\rhoin$, obtained from $\rho^0_{\rm coh}$ by setting the off-diagonal elements (coherences) to zero. We apply a generic unitary transformation described by $U_S(\Time)= e^{- i \vec{n} \cdot \vec{\sigma}}$, where $\vec{n} = \{n_x, n_y, n_z\}$ is a vector on the Bloch sphere that characterizes the system evolution.

To implement scheme A, we need to measure the relative phase between the two Fock states at the end of the evolution. In Fig.~\ref{fig:schemeA} (a,b) we plot the Q function of the initial (a) and final state of the cavity field (b). The phase change $\varphi$ undergone by the field, clearly observable in the phase space distribution, can be quantified by estimating the first moment of the field, that is, $\mean{a}$. Notice that $\varphi$ can only be determined as an ensemble average and not as the result of a single measurement.
The characteristic function $\GFunc_\lambda$ is obtained by repeating the characterization for different values of the coupling strength $\lambda$.
The real and imaginary part of $\GFunc_\lambda$ are plotted in Fig.~\ref{fig:schemeA} (c) for the initial state with and without coherences. While the real part (blue, solid) is the same for the two cases, the imaginary part (yellow, dashed and green, dot-dashed) is different, highlighting the contribution of the initial coherences. Recalling Eq.~\eqref{eq:moments}, we see that the two distributions differ already at the level of the first moment $\mean{W}=(-i) d \GFunc_\lambda/d \lambda|_{\lambda=0}$, that is, the average work performed on the system \cite{solinas2015fulldistribution}.
In Figure \ref{fig:schemeA} (d,e) we plot the Fourier transform of $\GFunc_\lambda$ for the initial states $\rhoin$ (d) and $\rhocoh$ (e), giving the quasiprobability distribution $\Prob(W)$. As $\Prob(W)$ is given by a sum of delta functions, it is displayed as a bar chart. The distribution corresponding to $\rhoin$ has peaks at the values $W=0, \pm \omega_a$. These peaks are immediately interpreted in terms of emission/absorption of an energy quantum with finite probability. The distribution corresponding to $\rhocoh$ looks identical except for two additional contributions at $W=\pm \omega_a/2$. These contributions, which cannot be explained in terms of classical probabilities, are equal in amplitude and opposite in sign. As a result, the distribution is normalized but not positive-definite, as anticipated in our discussion of Eq.~\eqref{eq:P(W)}.

\begin{figure}
    \begin{center}
    \includegraphics[width=\linewidth]{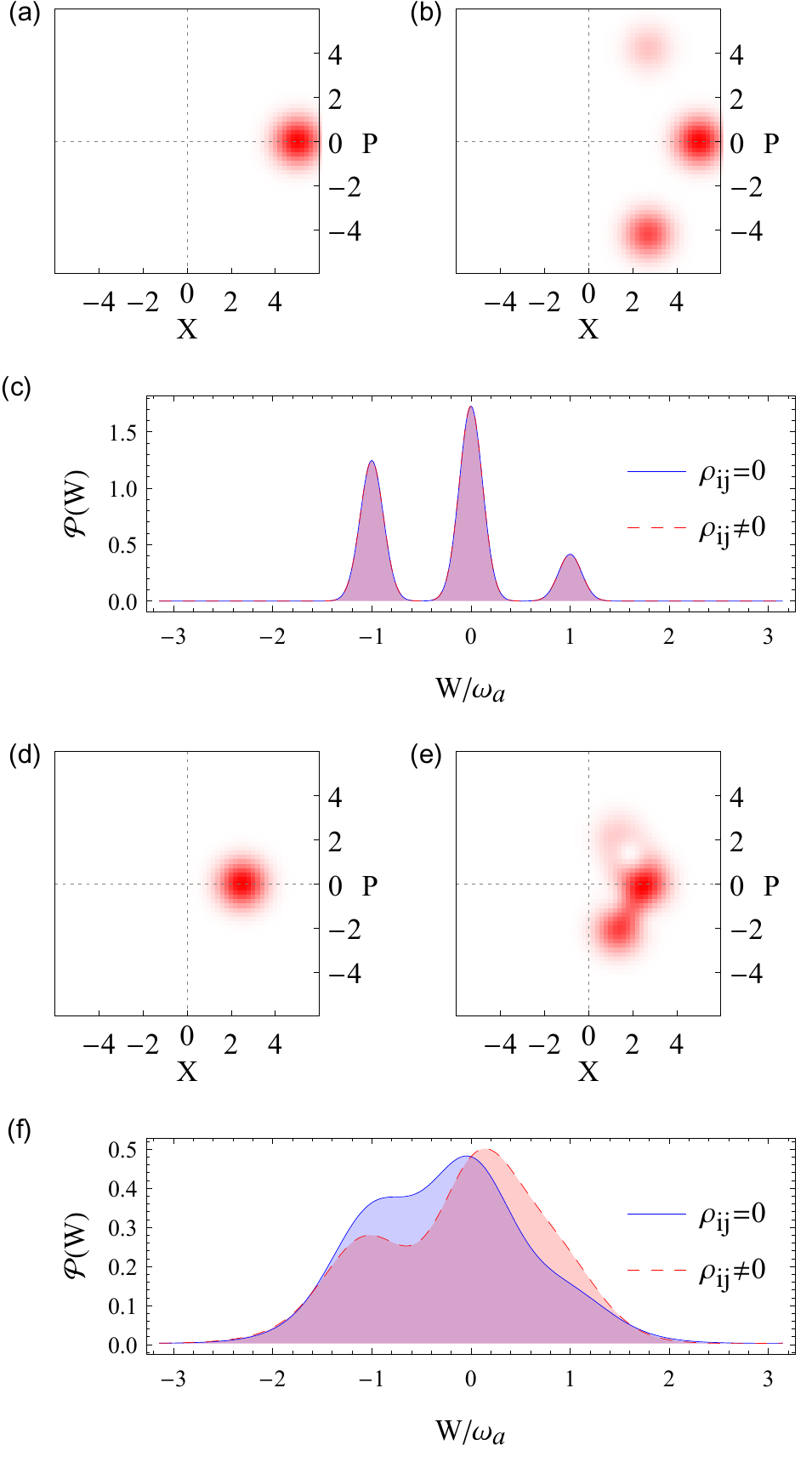}
   \end{center}
    \caption{\textbf{Results for scheme B.} (a,b) Initial (a) and final (b) Husimi-Q function for a detector prepared in a coherent state with $\alpha=5$. (c) Corresponding probability distribution, obtained by integrating the distribution (b) over the radial coordinate. The distribution is the same regardless of whether there is some initial coherence. (d,e) Same as in (a,b) for a coherent state with $\alpha=2.5$.
    (f) Corresponding probability distribution for an initial state without (blue, solid) and without initial coherence (red, dashed). The difference between the two distributions is purely due to quantum interference effects. 
    }
    \label{fig:schemeB}
\end{figure}

Scheme B can be implemented by initializing the field in a coherent state. The Q function of a coherent state, shown in Fig.~\ref{fig:schemeB} (a) for $\alpha=5$, is a Gaussian of variance 1, combining the uncertainty given by vacuum fluctuations ($1/2$) with the minimal uncertainty required by the uncertainty principle in order to perform simultaneous detection of both quadratures ($1/2$) \cite{Clerk2010}. Figure \ref{fig:schemeB} (b) shows the Q function of the corresponding final state of the field (the initial state of the qubit was $\rhoin$). The distribution consists of three Gaussians of different amplitude and same variance, whose mean position differs by a phase rotation. Each Gaussian corresponds to an admissible value for the work distribution, broadened by quantum fluctuations. Since the Gaussians are well separated in phase space, the outcome of an individual (``single-shot'') measurement can be assigned to a given value with high fidelity. As a matter of fact, the initial state $\rhocoh$ gives the same distribution as $\rhoin$, showing that initial coherences have no effect.
The phase-space distribution of Fig.~\ref{fig:schemeB} (b) can be mapped into the work distribution by projecting it on its angular component; the result is shown in Fig.~\ref{fig:schemeB} (c). Notice that the quantum uncertainty, represented by the width of the Gaussians, can be made negligible by choosing a coherent state with large $\alpha$, and/or by increasing the coupling.

As discussed previously, quantum interference effects can also be observed in scheme B, provided the measurement cannot fully distinguish between different outcomes. In the present setting, some degree of uncertainty is already embedded in the detector, as coherent states form an overcomplete basis, or, equivalently, they are not orthogonal.
The result of such a partial measurement is illustrated in Fig.~\ref{fig:schemeB} (d,e) for a coherent state of lower amplitude ($\alpha=2.5$), $\rhocoh$ as the initial state, and all other parameters the same as in Fig.~\ref{fig:schemeB} (a,b). The corresponding work distribution, shown in Fig.~\ref{fig:schemeB} (f) (red, solid), shows traces of quantum interference effects, as can be seen by comparing it against the distribution obtained from the incoherent initial state $\rhoin$ (blue, dashed). 
Notice that in the limit of a weak coherent state, $\alpha \ll 1$, the field state can be written as $\ket{\alpha}\approx \ket{0}+\alpha\ket{1}$ and the setting becomes the same as in scheme A (with reduced visibility). Therefore, the crossover between the two schemes can be continuously explored by changing the amplitude of the coherent state.

\section{Discussion}

Despite the fact that work cannot be associated to a hermitian operator, we have shown that it indeed presents typical quantum features, in that (i) its statistics depends on the measurement scheme used to retrieve it, (ii) it shows quantum interference patterns, (iii) it is influenced by quantum uncertainty, i.e., a revival of quantum interference is observed when we cannot distinguish between quantum trajectories leading to different energy exchanges.
The differences we observe in the work distribution boil down to the fact that in order to extract information from a quantum system we must let it interact with a detector and this interaction induces a backaction on the system dynamics \cite{talkner2016}. Introducing a quantum detector in our treatment \cite{solinas2015fulldistribution} makes it possible to unify and consistently interpret all these results.

In the first measurement scheme we considered (scheme A), the information is encoded in the phase accumulated by the detector during the coupling. This scheme has virtually no backaction; it can be used to retrieve all the moments of work, but at the price that the resulting distribution is not positive definite. While counter-intuitive, this reveals connections with a deeper and more fundamental aspect of quantum theory as the violation of a Leggett-Garg inequality \cite{Leggett1985, clerk2011full, Perarnau-Llobet2016}. Eventually, this apparent contradiction stems from the attempt of interpreting in a classical way a process that is intrinsically quantum \cite{mitchison2007sequential}. By contrast, the second, more conventional measurement scheme we introduced (scheme B) relies on a measurement of the detector position. In this case, the measurement backaction cannot be avoided but only reduced. While this approach gives a well-defined work probability distribution, it limits the amount of information that we can extract on the initial coherence of the system. 

To test these concepts, we have proposed an implementation in circuit QED. Using a cavity mode as the quantum detector, we have replaced momentum and position eigenstates by Fock states and coherent states, respectively. The reduced dimensionality of the Hibert space of the detector, as compared to the free-particle model, has no practical consequences for scheme A but introduces an intrinsic uncertainty in scheme B. This uncertainty can be understood as due to the overcompleteness of coherent states as a basis, or, equivalently, to the impossibility of measuring both field quadratures without introducing additional noise. Yet the relative uncertainty of the measurement outcome can be reduced to an arbitrary degree simply by increasing the amplitude of the coherent state, thereby recovering the result of a standard projective measurement. In this sense a single bosonic mode can be thought of as the minimal quantum detector capable of capturing the two complementary aspects of the work distribution discussed here. For comparison, an interferometric setup based on a single two-level system \cite{mazzola2013measuring,dorner2013extracting,Batalho2014} can be used to implement scheme A but does not possess enough logical space to encode the information to be extracted in scheme B.

A deeper understanding of the quantum features of work can lead to novel ways to exploit them. The natural question to ask is if and how these quantum effects can improve the efficiency in storing, manipulating and transporting energy at the atomic scale. Our results can already be used to estimate the energy cost of performing unitary operations on coherent quantum systems, for instance, in the context of quantum computation. The present work may be a starting point to investigate the role played by other quantum resources, such as entanglement, in the energy exchange between different quantum parties \cite{Perarnau-Llobet2015}, and to what extent the transport of heat, which is typically regarded as an incoherent form of energy, exhibits similar quantum effects. Altogether, these concepts may inspire a new generation of thermal machines exploiting quantum coherence at its full potential.

\begin{acknowledgments}
We gratefully acknowledge M.~Pechal, N.~Zangh\`i,  M. Perarnau-Llobet, A. Levy and R. Uzdin for fruitful discussions. 
P.S. has received funding from the European Union FP7/2007-2013 under REA
grant agreement no 630925 -- COHEAT and from MIUR-FIRB2013 -- Project Coca (Grant
No.~RBFR1379UX). S.G. acknowledges partial financial support from the Swiss National Science Foundation (SNF) Project 150046.
\end{acknowledgments}

%

\appendix
\newpage
\setcounter{equation}{0}

\begin{widetext}

\section{Supplemental Material}

\section{Density matrix in the momentum and position representation}
We consider a closed system for which the work done $W$ corresponds to the variation of the internal energy $\Delta U$.
The system is driven by a time-dependent classical external field and its Hamiltonian is $H_S(t)$.
The quantum detector is a free particle with position $x$ and momentum $p$ operators that it is coupled to the system by the interaction Hamiltonian $H_{SD}(t) =  -p ~ \beta(t) H_S(t) $, with a time-dependent coupling $\beta(t)$.
We take $\beta(t)= \lambda/p_0 [ \delta(t-\Time) - \delta(t)]$, where $\Time$ is the total evolution time, $\lambda$  is a time determining the strength of the coupling and $p_0$ is a rescaling parameter for the momentum.
The (system and detector) evolution operator is
\begin{equation}
 U(\Time)= e^{i    \frac{p \lambda}{p_0}  H_S(\Time) }  U_S(\Time)   e^{-i    \frac{p \lambda}{p_0}  H_S(0)}
 \label{eq_app:U_tot}
\end{equation}
with $U_S(\Time) = \Forder e^{-i \int_0^\Time dt H_S(t)}.$

We suppose that initially the system and the detector are not entangled and denote with $\{ \ket{\epsilon_i^t}\}$ and $\{ \ket{p}\}$ the basis  of $H_S(t)$ and the detector momentum, respectively.
The initial state reads  $\ket{\phi_0}  = \sum_i  \int d p~ \psi_i^0 G(p) \ket{\epsilon_i^0, p}$.
By applying the evolution operator (\ref{eq_app:U_tot}) to $\ket{\phi_0}$, we obtain 
\begin{eqnarray}
  \ket{\phi_0 } &\rightarrow& \sum_{i} \int dp \psi_i^0 G(p) e^{- i \frac{p \lambda \epsilon_i^0}{p_0} } \ket{\epsilon_i^0, p} 
  \rightarrow \sum_{ij} \int dp \psi_i^0 G(p) e^{- i \frac{p \lambda \epsilon_i^0}{p_0} } U_{S,ji} \ket{\epsilon_j^\Time, p} 
  \rightarrow  \sum_{ij} \int dp \psi_i^0 G(p) e^{i \frac{p \lambda \epsilon_{ji}}{p_0} } U_{S,ji} \ket{\epsilon_j^\Time, p} 
   \label{eq_app:phi_T}
\end{eqnarray}
where we have denoted with $\{ \ket{\epsilon_j^\Time}\}$ the eigenstates of $H_S(\Time)$, $U_{S,ji} = \matrixel{\epsilon_j^\Time}{U_S}{\epsilon_i^0}$ is the probability amplitude to go from $\ket{\epsilon_i^0}$ to $\ket{\epsilon_j^\Time}$ and $ \epsilon_{ji}= \epsilon_{j}^\Time -\epsilon_i^0$.
The corresponding final density matrix reads
\begin{equation}
\rho_\Time = \sum_{ijkj} \int  dp dp' \rho_{ik}^0  G(p) G^*(p') U_{S,ji} U_{S,kj}^\dagger e^{i \frac{p \lambda \epsilon_{ji}}{p_0}} e^{-i \frac{p'  \lambda \epsilon_{jk}}{p_0}} \ketbra{\epsilon_j^\Time , p}{\epsilon_l^\Time , p'}
 \label{eq_app:rho_tot_p}
\end{equation}
where $U_{S,kj}^\dagger= \matrixel{\epsilon_k^0}{U_S^\dagger}{\epsilon_j^\Time}$.
The work done, i.e., the internal energy variation $ \epsilon_{ji} $, is now encoded in the detector degrees of freedom.
Therefore, we focus on the detector degrees of  freedom by tracing out the system degrees of freedom, i.e., $\rho_D(\Time) = \Trace [\rho_\Time ] = \sum_l \matrixel{\epsilon_j^\Time }{\rho_\Time }{\epsilon_j^\Time }$.
In this way we obtain the detector density operator in the main text
\begin{equation}
\rho_D(\Time) = \sum_{ikj} \int  dp dp' \rho_{ik}^0 G(p) G^*(p') U_{S,ji} U_{S,kj}^\dagger e^{i \frac{p \lambda \epsilon_{ji}}{p_0}} e^{-i \frac{p' \lambda \epsilon_{jk}}{p_0}} \ketbra{ p}{p'}.
\label{eq:app_rho_p_basis}
\end{equation}

We can rewrite $\rho_D(\Time)$ in terms of the eigenstates $\ket{x}$ of the position operator $x$.
This can be done inserting the completeness operator $\int dx \ketbra{x}{x} = 1$ and using the relation $\braket{p|x} = e^{i x p}$.
We obtain 
\begin{equation}
 \rho_D(\Time) = \sum_{ikj} \int  dx dx' \rho_{ik}^0 g\left(x+\frac{\lambda \epsilon_{ji} }{p_0}\right) g^*\left(x'+ \frac{\lambda\epsilon_{jk}}{p_0}\right) U_{S,ji} U_{S,kj}^\dagger  \ketbra{x}{x'}
 \label{app_eq:rho_D_x}
\end{equation}
where $g(x) = \int d p \exp{\{ i p x \} } G(p)$ is the Fourier Transform of $G(p)$.
Thus, we have two operators and bases, i.e., the momentum $\{ \ket{p} \}$ and the position $\{ \ket{x} \}$ basis, to measure the detector and extract the information about $W$.

\section{Scheme A and B: Phase or Position Measurement}

We first discuss the details of the measurement scheme A.
The characteristic function $\GFunc_\lambda$ is related to the phase accumulated between two eigenstates of the momentum.
Recalling that the initial phase of the detector is $\matrixel{p_0/2 }{\rho^0_D }{-p_0/2 } = G(p_0/2) G^*(-p_0/2)$, the characteristic function is defined as $\GFunc_\lambda = \matrixel{p_0/2 }{\rho_D(t) }{-p_0/2 }/\matrixel{p_0/2 }{\rho^0_D }{-p_0/2 }$ \cite{solinas2015fulldistribution}. From Eq. (\ref{eq:app_rho_p_basis}), it explicitly reads 
\begin{equation}
  \GFunc_\lambda =\sum_{ikj}   \rho_{ik}^0 U_{S,ji} U_{S,kj}^\dagger \exp{\Big\{  i \lambda \Big[ \epsilon_j^\Time - \frac{( \epsilon_i^0+ \epsilon_k^0)}{2} \Big]\Big\}}.
\end{equation}
The (quasi-)probability distribution of the work (defined as $\Prob(W)= \int dp~\exp{\{ - i \lambda W \}} \GFunc_\lambda $) reads 
\begin{equation}
 \Prob(W) = \sum_{ikj}\rho_{ik}^0 U_{S,ji} U_{S,kj}^\dagger  \delta \Big[W -\left( \epsilon_j^\Time- \frac{\epsilon_i^0+\epsilon_k^0}{2} \right) \Big] =
 \sum_{ij} \rho_{ii}^0  |U_{S,ji}|^2  \delta \Big[W+ \epsilon_j^\Time- \epsilon_i^0\Big] + \sum_{i \neq k, j} \rho_{ik}^0  U_{S,ji} U_{S,kj}^\dagger \delta \Big[W-\left( \epsilon_j^\Time- \frac{\epsilon_i^0+\epsilon_k^0}{2} \right)\Big].
 \label{app_eq:P(W)}
\end{equation}
In the last equation we have separated the diagonal and off-diagonal matrix element contributions.
The first part can be directly interpreted in terms of (classical) transitions between energy eigenstates.
In this sense, it is a classical contributions.
On the contrary, the off-diagonal contributions in Eq. (\ref{app_eq:P(W)}) are purely quantum.
They lead to interference patterns in the work moments \cite{solinas2015fulldistribution} and cannot be interpreted in terms of classical transitions since they are associated to half-quantum energy exchanges.

Now we change perspective by using the detector as a standard pointer and by measuring its position after the interaction with the system (measurement scheme B). 
The ideal situation is when the detector is strongly localized in $x_0$: $g(x)=\delta(x-x_0)$.
The density matrix of the detector in Eq. (\ref{app_eq:rho_D_x}) reads 
\begin{equation}
  \rho_D(\Time) = \sum_{ikl} \int  dx dx' \rho_{ik}^0  \delta \left(x-x_0-\frac{\lambda \epsilon_{ji} }{p_0}\right) \delta \left(x'-x_0- \frac{\lambda \epsilon_{jk} }{p_0}\right) U_{S,ji} U_{S,kj}^\dagger  \ketbra{x}{x'}. 
\end{equation}
Notice that this results has an immediate interpretation and the detector acts as we expect.
It is initially in $x_0$ and it is shifted proportionally to the energy injected in (the work done on) the system.
Since it is a quantum detector it will be in a superposition of the position eigenstates, but the position measurement give us the work distribution.

The probability to measure $\bar{x}$ and, thus, to have a shift in the detector of $\Delta x= \bar{x}-x_0$ reads
\begin{equation}
  \Prob(\Delta x) =\matrixel{\bar{x}}{\rho_D}{\bar{x}}= \sum_{ikj} \rho_{ik}^0 U_{S,ji}  U_{S,kj}^\dagger \delta \left[ \frac{\lambda }{p_0} ( \epsilon_{i}^0 - \epsilon_{k}^0) \right]. 
  \label{app_eq:rho_T_gx}
\end{equation}

The work distribution is determined by the detector shift distribution of $\Delta x$.
We notice that $\Prob(\Delta x)\neq0$ only if  $\epsilon_{i}^0 - \epsilon_{k}^0$. For a non-degenerate system, this implies $i=k$ and we have $ \Prob(\Delta x) = \sum_{ij} |\psi_i^0|^2  |U_{S,ij}|^2 =   \sum_{ij} P_{i}^0 P_{i\rightarrow j}$.
where $P_{i}^0$ and  $P_{i\rightarrow j}$ are the probability to find the system initially in the state $\ket{\epsilon_i^0}$ and conditional probability to go from $\ket{\epsilon_i^0}$ to $\ket{\epsilon_j^\Time}$.
Therefore, the process can be interpreted in terms of classical probability and $\Prob(\Delta x)$ coincides with the classical contribution in Eq. (\ref{app_eq:P(W)}) and the one obtained with the the Two-Measurement Protocol \cite{kurchan2000quantum,tasaki2000jarzynski, esposito2009nonequilibrium, campisi2011colloquium}.

In realistic situations the detector cannot be not perfectly localized \cite{sokolovski2015meaning, talkner2016}.
The probability to have a shift $\Delta x = \bar{x} - x_0$ is given by Eq. (\ref{app_eq:rho_T_gx}) with the substitution $\delta(x) \rightarrow g(x)$.
Let us suppose that the detector state is described by a Gaussian function centered in $x_0$ with variance $\sigma$, i.e., 
$g(x)= \exp\{-\frac{(x-x_0 )^2}{4 \sigma ^2}  \}/ \sqrt[4]{2 \pi \sigma^2} $.
The presence of a finite variance implies that there is a uncertainty in the initial state of the detector and this affects our final measurement.
Analogously to Eq. (\ref{app_eq:rho_T_gx}), we obtain
\begin{eqnarray}
 \Prob(\Delta x) 
  &=& \sum_{ikj} \rho_{ik}^0 U_{S,ji} U_{S,kj}^\dagger
  \frac{e^{-\frac{\left(\Delta x- \frac{\lambda \epsilon_{ji} }{p_0}\right)^2 +\left(\Delta x- \frac{\lambda \epsilon_{jk} }{p_0}\right)^2 }{4 \sigma ^2}}}
  {\sqrt{2 \pi } \sigma }.
  \end{eqnarray}
In absence of initial coherences, i.e., $\rho_{ik}^0=0$, the system can be described by as a classical ensemble even if the resulting work distribution is blurred by the uncertainty.
The situation changes completely if coherences are present. 
In this case, the work distribution shows interference pattern as soon as there is an overlap between the Gaussian functions. We must to take into account the contributions depending on the off-diagonal terms of the initial density matrix $\rho_{ik}^0 $.
For a Gaussian distribution this occurs when  $\epsilon_k^0 - \epsilon_i^0 \approx  \sigma p_0/\lambda$.
Therefore, the effect of the interference becomes important when {\it the uncertainty is of the order of the energy gap we want to measure}.

\section{Circuit QED Hamiltonian}

The system-cavity Hamiltonian in the dispersive regime ($g/\Delta \ll 1$) reads $H' = H_{S} + H_c + H_I$ \cite{blais2004cavity, koch2007charge}
\begin{equation}
 H' = \Big( \omega_a + \frac{g^2}{\Delta}\Big) \sigma_z + \omega_r \hata^\dagger \hata + \frac{g^2}{\Delta} \sigma_z \hata^\dagger \hata.
 \label{eq_app:H_dispersive}
\end{equation} 
where $\omega_a$ and $\omega_r$ are the system and the cavity frequency, respectively, $\Delta = \omega_a - \omega_r$ is the controllable system-cavity detuning, $g$ is the system cavity coupling,
$\hata^\dagger$, $\hata$ and $\sigma_i$ are the creation and annihilation operators of the cavity and the Pauli operators acting on the system, respectively.

We consider the situation in which the system drive is periodic and occurring from time $0$ to time $\Time$.
Before and after the drive the system and the cavity are decoupled and $H_S(0)=H_S(\Time) = \omega_a \sigma_z$.
The detuning $\Delta$ can be controlled and becomes the time-dependent parameter to coupled and decouple the system and the cavity.
The time dependent Hamiltonian is
\begin{equation}
  H'(t)=
  \left \{ \begin{array}{ll}
  \omega_a  \sigma_z + \omega_r \hata^\dagger \hata  &~{\rm for~ } t<-\lambda \\
 \Big( \omega_a + \frac{g^2}{\Delta}\Big) \sigma_z + \omega_r \hata^\dagger \hata - \frac{g^2}{\Delta} \sigma_z \hata^\dagger \hata &~{\rm for~ }  -\lambda \leq t < 0 \\
 H_S(t) +  \omega_r \hata^\dagger \hata &~{\rm for~ } 0 \leq t \leq \Time \\
 \Big( \omega_a + \frac{g^2}{\Delta}\Big) \sigma_z + \omega_r \hata^\dagger \hata + \frac{g^2}{\Delta} \sigma_z \hata^\dagger \hata
 &~{\rm for~ } \Time <  t \leq  \Time +\lambda \\
   \omega_a  \sigma_z + \omega_r \hata^\dagger \hata &~{\rm for~ } t >\Time +\lambda
\end{array} 
\right .\
\end{equation}
In other terms, initially we couple the system and the detector bringing $\omega_a$ closer to the resonance for  time $\lambda$.
Then we decouple them and act only on the system with an arbitrary transformation (associated to the time-dependent Hamiltonian $H_S(t)$ for $0 < t < \Time$).
Finally, we couple the system and the detector again bringing $\omega_a$ closer to the resonance for  time $\lambda$.
Notice that in the first coupling $\Delta$ must be negative, i.e., $\omega_a < \omega_r$, while in the second one must be positive, i.e., $\omega_a > \omega_r$. 
The total unitary operator is \cite{solinas2015fulldistribution}
\begin{equation}
 U = e^{- i \frac{\lambda  g^2}{\Delta} \sigma_z  \hata^\dagger \hata} U_S e^{ i \frac{\lambda  g^2}{\Delta} \sigma_z \hata^\dagger \hata}.
 \label{app_eq:U_tot}
\end{equation}

We denote with $\ket{\eta_i}$ ($\eta_i=\pm1$) and $\ket{n}$ ($n$ integer) the eigenstates of the system $\sigma_z$ operator and Fock state of the cavity Hamiltonian, respectively. 
Taking the initial state to be $\ket{\phi_0 } = \sum_{i, n} \psi_i^0 G_n^0  \ket{\eta_i, n}$, the sequence of applied operators from Eq. (\ref{app_eq:U_tot}) gives
\begin{eqnarray}
  \ket{\phi_0 } &\rightarrow& \sum_{i, n} \psi_i^0 G_n^0 e^{- i \lambda \left(\omega_r - \frac{g^2}{\Delta} \eta_i \right) n} \ket{\eta_i, n} \rightarrow  \sum_{i, n} \psi_i^0 G_n^0 e^{- i \lambda \left(\omega_r - \frac{g^2}{\Delta}\eta_i \right) n}  U_S \ket{\eta_i, n} \rightarrow \sum_{i, n, j} \psi_i^0 G_n^0 e^{- 2 i \lambda \omega_r} e^{- i  \frac{\lambda g^2}{\Delta} (\eta_j-\eta_i ) n}  U_{S,ji} \ket{\eta_j, n}.
   \label{eq_app:phi_dispersive}
\end{eqnarray}
Here, we have also kept track of the dynamics phase accumulated by the detector. In circuit-QED there contribution can be eliminated with opportune transformation.
Therefore, in the following we neglect them.
By extending this result to a general initial density matrix and then tracing out the system degrees of freedom we obtain 
\begin{equation}
\rho_D(\Time) = \sum_{ikj} \rho_{ik}^0 G_n^0 (G_m^0)^* U_{S,ji} U_{S,kj}^\dagger e^{i\frac{ \lambda g^2}{\Delta} [\eta_j(n-m) +\eta_i n -\eta_k m ]}  \ketbra{ n}{m}
\label{eq_app:rho_cavity}
\end{equation}

\subsection{Scheme A: phase measurement}

The generating function is obtained by the phase accumulated between the eigenstates of the cavity $ \matrixel{\bar{n} + \delta n}{ \rho_D(\Time)}{ \bar{n} -\delta n }/ \matrixel{\bar{n}+ \delta n }{ \rho_D(0)}{ \bar{n}- \delta n}$ where $\rho_D(t) = \Trace_S[ \rho(t)]$.
The most convenient choice is $\delta n =1$ and, noticing that $\matrixel{n}{ \rho_D(0)}{ m} = G_n^0 (G_m^0)^*$, we obtain 
\begin{eqnarray}
 \GFunc_\lambda &=& \sum_{ikj} \rho_{ik}^0 U_{S,ji} U_{S,jk}^\dagger e^{ i \frac{2 \lambda g^2}{\Delta}  ( \eta_j- \frac{\eta_i+\eta_k}{2} ) }
\end{eqnarray} 
where we have neglected a contribution $ \bar{n} (\eta_i - \eta_k)$ that does not contain information about the final energy $\eta_j$.
Notice that in this case the measure output is twice the work done but this can easily taken into account in the data analysis.
The (quasi-)probability distribution of the work is obtained by $\Prob(W) = \int d \lambda \exp\{{- i W \lambda}\}  \GFunc_\lambda$.

\subsection{Scheme B: Direct measure of the the work distribution}

To measure directly the work distribution in a cQED system, we use the coherent states \cite{gerry2005introductory}.
In fact, the system-cavity Hamiltonian produces a transition between coherent state that is proportional to the energy of the system.

The initial state is now taken as  $\ket{\phi_0 } =  \sum_{i} \psi_i^0 \ket{\eta_i, \alpha}$ where $\ket{\alpha}$ is a coherent state defined as $\ket{\alpha} =  D(\alpha)  \ket{0} =  e^{\alpha (\hata^\dagger -  \hata)}  \ket{0}$ and $D(\alpha)$ is the displacement operator.
In the following we take $\alpha$ real.
The $\alpha$ parameter is directly related to the average number of photons and their variance by the relations $|\alpha| = \sqrt{\average{\hata^\dagger \hata}}= \sqrt{\average{n}}$ and $|\alpha| = \average{\Delta n} $.

The dynamics of the coherent state can be calculated as in Eq. (\ref{eq_app:phi_dispersive}) and  using the formula  $e^{\pm i \chi \hata^\dagger a} \ket{\alpha} = \ket{\alpha e^{\pm i \chi}}$ \cite{gerry2005introductory}.
The effect of the system-cavity interaction is particularly interesting in the phase space \cite{gerry2005introductory}.
The average position and momentum of the quantum oscillator in a coherent state $\ket{\bar{\alpha}}$ are $\average{X}= 1/2 (\bar{\alpha}+\bar{\alpha}^*)$ and $\average{P}= -i/2 (\bar{\alpha}-\bar{\alpha}^*)$ \cite{gerry2005introductory}.
If we take $\bar{\alpha}=\alpha$ real, the initial position in the phase space is $(\average{X}_0, \average{P}_0) = (\alpha,0)$.
After the interaction the state is in a superposition of different coherent state.
For any energy difference $\epsilon $, we have a contribution in position $(\average{X}_\Time, \average{P}_\Time) = (\alpha \cos \epsilon, \alpha \sin \epsilon)$.
The measure of the position-momentum in the phase space of the cavity allows us to extract information about the angle $\epsilon$ and, thus, the exchanged energy.

Here, instead of recalculating the dynamics for the coherent state, we can use the cavity density matrix obtained by Eq. (\ref{eq_app:rho_cavity}).
We write $\xi = \delta e^{i \theta}$.
The matrix element $ \Prob(\xi) = \matrixel{\xi}{\rho_D}{\xi}$ gives us the probability to measure the coherent state $\ket{\xi}$. It can be calculated using the formula $\braket{m|\xi} = \exp{(-|\xi|^2/2)}\xi^m/\sqrt{m!}$ \cite{gerry2005introductory}.
The initial state projections are $G_n^0 = \braket{n|\alpha} = \exp{(-\alpha^2/2)}\alpha^n/\sqrt{n!}$ and, since $\alpha$ is real, $(G_m^0)^* = G_m^0$. 
By collecting the sum over $n$ and $m$ and using $\sum_{n=0}^\infty (\xi^* \alpha)/n! = \exp{(\xi^* \alpha)}$, we obtain 
\begin{eqnarray}
Q(\xi)&=& \frac{1}{\pi} \sum_{p,i,k} \rho_{ik}^0 U_{S,ji} U^\dagger_{S,kp} 
 e^{-\delta^2 -\alpha^2 + \alpha \delta \chi_{jik}}  = \frac{1}{\pi} \sum_{ikj} \rho_{ik}^0 U_{S,ji} U^\dagger_{S,kj} e^{-\Big(\delta -\frac{\alpha \chi_{jik}}{2}\Big)^2} e^{-\alpha^2 \Big(1-\frac{\chi_{jik}^2}{4}\Big)} 
   \label{eq_app:P_xi_distribution}
\end{eqnarray}
where $\chi_{jik} = e^{-i \left(\Lambda_{ji}+\theta \right)} + e^{i \left(\Lambda_{jk}+\theta \right)}$, $\Lambda_{ji} = g^2 \lambda  (\eta_j - \eta_i)/\Delta = g^2 \lambda  \eta_{ji}/\Delta$ and $Q(\xi) =  \Prob(\xi)/\pi$ is the Husimi function \cite{gerry2005introductory}.

The information about the quantum work is encoded only in the angle $\theta$ of the coherent state.
For this reason, it is convenient to integrate the 'radial' contribution $\delta$ to obtain $\Prob(W) = \Prob(\theta) = \int_0^\infty d \delta~ \delta~Q(\xi)$
(notice that we $\delta$ and $\theta$ are polar coordinated and the integration must be with $d \delta~ \delta$ \cite{gerry2005introductory}).
We obtain 
\begin{eqnarray}
 \Prob(W) = \Prob(\theta) &=&\sum_{ijk} \rho_{ik}^0 U_{S,ji} U^\dagger_{S,kj} \Big\{ \frac{e^{-\alpha^2}}{2 \pi} +
 \frac{e^{-\alpha^2 (1-\frac{\chi_{jik}^2}{4})}}{4 \sqrt{\pi}}\alpha \chi_{jik} \Big[1+ {\rm Erf}\Big(\frac{\alpha \chi_{jik}}{2}\Big) \Big] \Big\}
 \label{eq_app:Prob_cQED}
\end{eqnarray} 
where ${\rm Erf}(x)$ is the error function.
Equation (\ref{eq_app:Prob_cQED}) gives us the measurable work distribution as a function of $\theta$ {\it and} the initial state detector $\alpha$.
Below we will see that $\alpha$ sets of the detector accuracy and, therefore, plays a pivotal role in the revival of quantum interference effect.

For $\alpha> 1$, the first term in curly brakets in Eq. (\ref{eq_app:Prob_cQED}) is exponentially suppressed.
Since $|{\rm Erf}(x)| \leq 1$, the behaviour of the remainin contributions is determined by the exponent $\exp\{-\alpha^2 (1-\frac{\chi_{jik}^2}{4})\}$.
We expect that the dominant contributions are the ones for which 
\begin{equation}
 \alpha^2 \left(1-\frac{\chi_{jik}^2}{4} \right) \leq 1
 \label{app_eq:alpha_condition}
\end{equation}
since the other contributions are exponentially suppressed.

We can consider $\theta$ centered around $ g^2 \lambda  \eta_{ji}/\Delta$ associated to $\Lambda_{jk}$ and take as variable the distance $\Delta \theta $ from this point: $\theta = - \Lambda_{jk} + \Delta \theta$.
We split the contributions in the exponent in the real and imaginary part and focus on the first one: $ \Real (\chi_{jik}^2)= 4 \cos \left(\Lambda_{ik}\right) \cos ^2\left(\Delta \theta -\frac{\Lambda_{ik}}{2}\right)$.

The diagonal term in Eq. (\ref{eq_app:Prob_cQED}) are found for $k=i$.
In these cases, $\Real (\chi_{jii}^2) = 4  \cos ^2(\Delta \theta)$ and the condition (\ref{app_eq:alpha_condition}) reads $2 \alpha^2 \sin^2 (\Delta \theta/2) \leq 1$.
Thus, the main contributions are found for $\Delta \theta \approx 2 /\alpha$, i.e., localized around to $- \Lambda_{jk}$.  Away from this values we have an exponential suppression.

Now we focus on the off-diagonal terms in Eq. (\ref{eq_app:Prob_cQED}) are found for $k\neq i$.
The solutions of the equation $\alpha^2 (1-\frac{\Real (\chi_{jik}^2)}{4}) \approx 1$  are of the form
\begin{equation}
 \Delta \theta = \arccos \left(\sqrt{ \frac{1-1/\alpha^2}{ \cos \Lambda_{ik}}} \right)+\frac{\Lambda_{ik}}{2}.
\end{equation} 
The condition for this to be real is $\cos \Lambda_{ik} \leq 0$.
The argument of the $\arccos$ must be $-1 \leq (1-1/\alpha^2)/ \cos \Lambda_{ik}\leq 1$.
For large $\alpha$ we must consider only the right inequality that reads $\cos \Lambda_{ik} \leq 1- 1/\alpha^2$.
The latter can be satisfied only for small $\Lambda_{ik}$ and by expanding the cosine function we arrive at 
\begin{equation}
 \Lambda_{ik}  = \lambda g^2 \lambda  \eta_{ik}/\Delta \leq \frac{\sqrt{2}}{\alpha}.
 \label{app_eq:interference_condition}
\end{equation}

If condition (\ref{app_eq:interference_condition}) is not satisfied the only non-vanishing contributions are those coming from the diagonal terms.
The quantum work probability distribution reads $\Prob(W) = \sum_{ij} |\psi_i^0|^2  U_{S,ji} U^\dagger_{S,ij} =  \sum_{ij} P_{i}^0 P_{i \rightarrow j}$ where $P_{i}^0 = \rho_{ii}^0$ is the probability to find the system initially in the state $i$ and $P_{i \rightarrow j} = |U_{S,ji}|^2$ is the probability to go from $i$ to $p$.
On the contrary, if condition (\ref{app_eq:interference_condition}) is satisfied we must keep all the contributions including the one depending on the off-diagonal initial density matrix.

Therefore, in order for these terms to be relevant, the normalized energy exchange $\eta_{ik}$ that we aim to measure must be smaller then the average number of photons in the coherent state.
In other words, the initial coherent state of the cavity sets the limit for the measurement precision.
Below the limit (\ref{app_eq:interference_condition}) we cannot resolve the (normalized) energy variation of $ \eta_{ik}$ and interference effects change the work distribution.

%

\end{widetext}


\end{document}